\documentclass[prl,aps,showpacs,twocolumn]{revtex4}%
\usepackage{graphicx}
\usepackage{amsmath}
\usepackage{amsfonts}
\usepackage{amssymb}%
\providecommand{\U}[1]{\protect\rule{.1in}{.1in}}
\begin{document}
\title{Faddeev-Jackiw Quantization of Non-Autonomous Singular Systems }
\author{Zahir Belhadi$^{a,b}$, Alain B\'{e}rard$^{b}$\ and Herv\'{e} Mohrbach$^{b}$}
\affiliation{$a$\ - Laboratoire de physique th\'{e}orique, Facult\'{e} des sciences
exactes, Universit\'{e} de Bejaia. 06000 Bejaia, Alg\'{e}rie}
\affiliation{$b$\ - Equipe BioPhyStat, ICPMB, IF\ CNRS\ N$%
{{}^\circ}%
2843$ Universit\'{e} de Lorraine, 57070 Metz Cedex, France.}

\begin{abstract}
We extend the quantization \`{a} la Faddeev-Jackiw for non-autonomous singular
systems. This leads to a generalization of the Schr\"{o}dinger equation for
those systems. The method is exemplified by the quantization of the damped
harmonic oscillator and the relativistic particle in an external
electromagnetic field.

\end{abstract}
\maketitle

The quantization of constrained systems is almost as old as the beginning of
quantum mechanics. It was Dirac \cite{dirac} who elaborated a Hamiltonian
approach with a categorization of constraints and the introduction of the
so-called Dirac brackets. Later, Faddeev and Jackiw \cite{FJ} suggested an
alternative and generally simpler method based on a symplectic structure.
Recently, we have proposed a third approach for classically soluble
constrained\textit{ }systems where the brackets between the constants of
integration are computed. This method does neither require Dirac formalism nor
the symplectic method of Faddeev-Jackiw \cite{nous}. All three approaches were
developed for autonomous constrained systems only. The quantization of
non-autonomous singular systems has turned to be non trivial \cite{mukunda}.
Gitman and Tyutin \cite{GT}, via notably the introduction of a conjugate
momentum of time, could extend Dirac approach and brackets for those systems.
In the present work, our aim is to generalize the Faddeev-Jackiw symplectic
approach to non-autonomous constrained systems. This leads to a generalization
of the Schr\"{o}dinger equation which encompasses non-autonomous singular
systems. The quantization of a relativistic particle in an electromagnetic
field is solved by this method, constituting an original derivation of the
Dirac equation.

Consider a non autonomous Lagrangian one-form as in \cite{FJ}%
\begin{equation}
Ldt=a_{j}(\xi,t)d\xi^{j}-H(\xi,t)dt, \label{L1}%
\end{equation}
where $\xi^{j}$ are the $N$-component phase-space coordinates with
$j=1,...,N$, where $a_{j}$ and $H$ are time-dependent. The Euler-Lagrange
equations lead to%
\begin{align}
\dot{\xi}^{i}  &  =f_{ij}^{-1}\left(  \frac{\partial H}{\partial\xi^{j}}%
+\frac{\partial a_{j}}{\partial t}\right) \label{EL}\\
f_{ij}  &  =\frac{\partial a_{j}}{\partial\xi^{i}}-\frac{\partial a_{i}%
}{\partial\xi^{j}} \label{FiJ}%
\end{align}
for an invertible antisymmetric matrix $f_{ij}$. The dot denotes
differentiation with respect to $t.$ Eq.\ $\left(  \ref{EL}\right)  $ can not
be derived from Hamilton equations through brackets when $\partial
a_{j}/\partial t\neq0$ (see Eq. $\left(  \ref{wrong}\right)  $). Canonical
quantization seems compromised in this case.\ To solve this problem via the
Faddeev-Jackiw approach we first introduce a time parameter $\tau$ such that
time is promoted to a dynamically variable $t=t(\tau)$ and $\xi^{i}=\xi
^{i}(\tau)$. This leads to a Lagrangian $L_{\tau}$, given by $L_{\tau}%
d\tau=Ldt$, so that the action remains the same. $L_{\tau}$ has a gauge
invariance due to the arbitrariness of the parameter $\tau$. We thus define a
new Lagrangian $\tilde{L}_{\tau}=L_{\tau}+\omega\left(  t^{\prime}-1\right)  $
that implements the gauge constraint $t^{\prime}=1$ (prime denotes
differentiation with respect to $\tau$) via $\omega(\tau)$ a Lagrange
multiplier seen as a new variable. Dropping a total time derivative term, one
ends up with the following Lagrangian one-form%
\begin{equation}
\widetilde{L}d\tau=a_{j}(\xi,t)d\xi^{j}-H(\xi,t)dt-td\omega-\widetilde{H}%
d\tau, \label{L2}%
\end{equation}
where $\widetilde{H}(\xi,t,\omega)=\omega$ defines a new Hamiltonian.\ The
Euler-Lagrange equations with $\widetilde{L}$ lead to Eq $\left(
\ref{EL}\right)  $ and the equations $\omega^{\prime}=0$ and $t^{\prime}=1$.
Unlike the initial Lagrangian $\left(  \ref{L1}\right)  $, $\widetilde{L}$ is
autonomous with respect to $\tau$ and Eq. $\left(  \ref{L2}\right)  $ is
precisely of the form studied by\ Faddeev-Jackiw \cite{FJ}%
\begin{equation}
\widetilde{L}d\tau=A_{i}(\zeta)d\zeta^{i}-\widetilde{H}(\zeta)d\tau,
\label{FJL}%
\end{equation}
with now $\zeta^{i}$ the $N+2$-component phase space coordinates defined by
$\zeta^{i}=\xi^{i}$ for $i=1...N$; $\ \zeta^{N+1}=t$ and $\zeta^{N+2}%
=\omega.\ $Here $A_{i}=a_{i}$ for $i=1...N$ and $A_{N+1}=-H,$ $A_{N+2}=-t.\ $

Following \cite{FJ} we first consider the Euler-Lagrange equations%
\begin{equation}
F_{ij}\zeta^{j\prime}=\frac{\partial\widetilde{H}}{\partial\zeta^{i}}
\label{ELFJ}%
\end{equation}
with%
\begin{equation}
F_{ij}=\frac{\partial A_{j}}{\partial\zeta^{i}}-\frac{\partial A_{i}}%
{\partial\zeta^{j}}. \label{FJdef}%
\end{equation}
If the antisymmetric matrix $F_{ij}$ is regular then Eq. $\left(
\ref{ELFJ}\right)  $ becomes%
\begin{equation}
\zeta^{i\prime}=F_{ij}^{-1}\frac{\partial\widetilde{H}}{\partial\zeta^{j}}.
\end{equation}
Writing the Hamilton equations%
\begin{equation}
\zeta^{i\prime}=\left\{  \zeta^{i},\widetilde{H}\right\}  =\left\{  \zeta
^{i},\zeta^{j}\right\}  \frac{\partial\widetilde{H}}{\partial\zeta^{j}}
\label{H1}%
\end{equation}
the generalized brackets for non-autonomous systems are readily obtained as%
\begin{equation}
\left\{  \zeta^{i},\zeta^{j}\right\}  =F_{ij}^{-1}\text{ \ \ \ \ }i,j=1...N+2.
\label{bracket}%
\end{equation}
Now from the matrix $F_{ij}^{-1}$ we obtain the following fundamental brackets%
\begin{align}
\left\{  \xi^{i},\xi^{j}\right\}   &  =f_{ij}^{-1}\text{ \ \ \ \ \ }%
i,j=1...N\label{C1}\\
\left\{  \xi^{i},t\right\}   &  =0,\label{C2}\\
\left\{  t,\omega\right\}   &  =1,\label{C3}\\
\left\{  \xi^{i},\omega\right\}   &  =f_{ij}^{-1}\left(  \frac{\partial a_{j}%
}{\partial t}+\frac{\partial H}{\partial\xi^{j}}\right)  . \label{C4}%
\end{align}
The brackets $\left\{  \xi^{i},\xi^{j}\right\}  =f_{ij}^{-1}$ which are
unchanged with respect to the autonomous case are sufficient to deduce all
brackets of the theory. We see that $\omega$ plays the role of a conserved
Hamiltonian for the time evolution of $\xi^{i}$. Indeed, let us now use the
fact that $\widetilde{H}(\xi,t,\omega)=\omega.$ Eq.$\left(  \ref{H1}\right)  $
then reduces to the "Hamilton" equations of motion%

\begin{equation}
\frac{d\zeta^{i}}{d\tau}=\left\{  \zeta^{i},\omega\right\}  \text{
\ \ \ \ \ }i=1...N+2. \label{EqH}%
\end{equation}
For $i=N+2$ and $N+1,$ Eq. $\left(  \ref{EqH}\right)  $ gives respectively
$d\omega/d\tau=0$ so $\omega$ is a constant and $dt/d\tau=1,$ so $t(\tau
)=\tau$ without loss of generality$.$ The dynamics of the phase-space
coordinates $\xi^{i}(t)$ with respect to the physical time $t$ is therefore%
\begin{equation}
\dot{\xi}^{i}=\left\{  \xi^{i},\omega\right\}  =f_{ij}^{-1}\frac{\partial
H}{\partial\xi^{j}}+f_{ij}^{-1}\frac{\partial a_{j}}{\partial t}, \label{EqFJ}%
\end{equation}
for $i=1..N.\ $These equations are the same as the Euler-Lagrange equations
Eq.$\ \left(  \ref{EL}\right)  $. Eq. $\left(  \ref{EqFJ}\right)  $ is the
generalization of the Hamilton equation for a non-autonomous system. Indeed
the equations of motion can not be obtained by usual Hamilton equations which
would give instead%
\begin{equation}
\dot{\xi}^{i}=\left\{  \xi^{i},H\right\}  =f_{ij}^{-1}\frac{\partial
H}{\partial\xi^{j}} \label{wrong}%
\end{equation}
correct for the case $\partial a_{j}/\partial t=0$ only.

The time derivative of any function $O(\xi^{i},t)=\dot{\xi}^{j}\partial
O/\partial\xi^{j}+\partial O/\partial t$ is also obtained as%
\begin{equation}
\dot{O}(\xi^{i},t)=\left\{  O,\omega\right\}  . \label{dg}%
\end{equation}
This is a new general relation that embraces all systems, regular and
autonomous as well. It does not contain the term $\partial O/\partial t$. The
Hamiltonian $H$ is generally not conserved for a non-autonomous system
contrary to $\omega$ and $\dot{H}=\left\{  H,\omega\right\}  =-\dot{\xi}%
^{j}\partial a_{j}/\partial t+\partial H/\partial t$ $\left(  =-\partial
L/\partial t\right)  $.

Before considering the quantization, we note from Eq. $\left(  \ref{ELFJ}%
\right)  $ that the condition of regularity of the matrix $F_{ij}$ is actually
reduced to the regularity of $f_{ij}.$ Now, for a singular matrix $f_{ij}$ we
recall briefly the procedure which consists first in determining the
zero-modes $v^{(\alpha)},$ which are solutions of $v_{i}^{(\alpha)}f_{ij}=0.$
Multiplying Eq. $\left(  \ref{EqFJ}\right)  $ by $v_{i}^{(\alpha)}$ we obtain
the equation $v_{i}^{(\alpha)}\left(  \partial H/\partial\xi_{i}+\partial
a_{i}/\partial t\right)  =0$ where we used the equations $\partial
\widetilde{H}/\partial\omega=1$ and $\partial\widetilde{H}/\partial\xi
_{i}=\partial\widetilde{H}/\partial t=0$. These relations between the
variables $\xi_{i}$ and $t$ constitute a set of constraints $\phi_{\alpha}%
(\xi,t)=0$ that must be conserved in time $d\phi_{\alpha}/d\tau=0.$ We
therefore add to the Lagrangian $\widetilde{L}$ the term $\lambda_{\alpha
}^{\prime}\phi_{\alpha}$ to obtain the new autonomous Lagrangian
$\widetilde{L}+\lambda_{\alpha}^{\prime}\phi_{\alpha},$ and redo the
Faddeev-Jackiw procedure considering $\lambda_{\alpha}$ as new independent
variables. Now, if the new matrix $f_{ij}$\ is invertible, all brackets are
accessible. If not we calculate the zero-modes that will give in principle new
constraints that we should add to the Lagrangian and so on. At the end, either
we obtain an invertible matrix and we can determine the brackets, or the
matrix is always singular with no new constraints. In this case, our initial
Lagrangian has gauge symmetry that should be fixed by using additional
conditions to obtain a non singular matrix $f_{ij}$.

\bigskip

\textit{Quantization. }We first remark that from Eqs. $\left(  \ref{EqFJ}%
\right)  $ and $\left(  \ref{wrong}\right)  $ $\omega$ can be decomposed as
$\omega=\varepsilon+H$ with the brackets%
\begin{equation}
\left\{  \xi^{i},H\right\}  =f_{ij}^{-1}\frac{\partial H}{\partial\xi^{j}%
}\text{ \ \ and \ \ }\left\{  \xi^{i},\varepsilon\right\}  =f_{ij}^{-1}%
\frac{\partial a_{j}}{\partial t}. \label{eps}%
\end{equation}
Eqs. $\left(  \ref{C2}\right)  $ and $\left(  \ref{C3}\right)  $ show that
\ $\left\{  t,\varepsilon\right\}  =1$ so $\varepsilon$ is a variable
conjugate to time. It was introduced in \cite{GT} for singular systems and in
\cite{Edelen} for regular ones. Although it is not suitable to talk about
momentum conjugate in the Faddeev-Jackiw approach it is nevertheless
interesting to see that $\varepsilon$ is actually equal to the conjugate
momentum of time defined as $p_{t}(t)=(\partial\tilde{L}_{\tau}/\partial
t^{\prime})_{\tau=t}.$ Thus $\omega=p_{t}+H(p,q,t)$ can be interpreted as an
extended Hamiltonian in an extended phase space \cite{Rovelli}.

For the quantization in the Schr\"{o}dinger picture we define the operators
$\hat{\zeta}^{i}$ associated to their classical counterparts. Any operator
$\hat{O}$ associated to a classical function $O(\zeta)$ is defined by the rule
$\hat{O}=O(\hat{\zeta})$ and commutators are defined as $\left[  ,\right]
=i\hbar\left\{  ,\right\}  _{\zeta=\hat{\zeta}}$ (we disregard problems with
operator ordering). Therefore from Eq. $\left(  \ref{bracket}\right)  $ we
have%
\begin{equation}
\left[  \hat{\zeta}^{i},\hat{\zeta}^{j}\right]  =i\hbar\left\{  \zeta
^{i},\zeta^{j}\right\}  _{\zeta=\hat{\zeta}}=i\hbar F_{ij}^{-1}(\hat{\zeta}),
\label{Q}%
\end{equation}
thus in general for singular systems operators do not satisfy the canonical
commutation relations. An operator $\widehat{\dot{O}}$ associated to $\dot
{O}(\xi,t)$ will be given by the quantum version of Eq. $\left(
\ref{dg}\right)  $%
\begin{equation}
\widehat{\dot{O}}=\frac{1}{i\hbar}[\hat{O},\hat{\omega}]=\left\{
O,\omega\right\}  _{\zeta=\hat{\zeta}},
\end{equation}
where $\hat{\omega}=\hat{\varepsilon}+\hat{H}.$ As $[\hat{t},\hat{\varepsilon
}]=i\hbar$, we define naturally $\hat{t}=t$ and thus $\hat{\varepsilon
}=-i\hslash d/dt$ is the time translation operator \footnote[1]{Rigorously
$\hat{\varepsilon}=-i\hbar d/dt+\hat{\sigma}$ with $[\hat{t},\hat{\sigma}]=0.$
But in this case Eq.\ $\left(  \ref{GSschro}\right)  $ becomes $i\hbar\frac
{d}{dt}\left\vert \psi\right\rangle =\hat{H}_{eff}\left\vert \psi
(t)\right\rangle $ with $\hat{H}_{eff}=\hat{H}+\hat{\sigma}$.
Additionally\ Eq. $\left(  \ref{de1}\right)  $ becomes \ $d\hat{\xi}%
^{i}/dt=(f_{ij}^{-1}\frac{\partial a_{j}}{\partial t})_{\xi=\hat{\xi}}%
-[\hat{\xi}^{i},\hat{\sigma}]$. Since for $\partial a_{j}/\partial t=0$ we
have $d\hat{\xi}^{i}/dt=0$ in the Schr\"{o}dinger picture, we put $\hat
{\sigma}=0$.}.\ Therefore $d\hat{O}/dt=\frac{1}{i\hbar}[\hat{O},\hat
{\varepsilon}]$ and $\widehat{\dot{O}}$ can also be written%
\begin{equation}
\widehat{\dot{O}}=\frac{1}{i\hbar}[\hat{O},\hat{H}]+\frac{d\hat{O}}{dt},
\end{equation}
which is different from the usual expression $\widehat{\dot{O}}=\frac
{1}{i\hbar}[\hat{O},\hat{H}]+\partial\hat{O}/\partial t.$ The reason is that
$\hat{\xi}^{i}$ is explictly time dependent and its evolution is given by%
\begin{equation}
\frac{d}{dt}\hat{\xi}^{i}=\left\{  \xi^{i},\varepsilon\right\}  _{\xi=\hat
{\xi}}=\left(  f_{ij}^{-1}\frac{\partial a_{j}}{\partial t}\right)  _{\xi
=\hat{\xi}}. \label{de1}%
\end{equation}
The solution of this differential equation gives us $\hat{\xi}^{i}(t)$ with
$\hat{\xi}^{i}(0)=\hat{\xi}_{s}^{i}$ the usual time independant
Schr\"{o}dinger operator satisfying $[\hat{\xi}_{s}^{i},\hat{\xi}_{s}%
^{j}]=i\hbar f_{ij}^{-1}(\hat{\xi}_{s})$ (note that this quantization at $t=0$
could be preformed at any arbitrary time \cite{GT}). The operator
$\widehat{\dot{\xi}}^{i}$ associated to $\dot{\xi}^{i}$ is therefore%
\begin{equation}
\widehat{\dot{\xi}}^{i}=\left\{  \xi^{i},H+\varepsilon\right\}  _{\xi=\hat
{\xi}}=\left(  f_{ij}^{-1}\frac{\partial H}{\partial\xi^{j}}+f_{ij}^{-1}%
\frac{\partial a_{j}}{\partial t}\right)  _{\xi=\hat{\xi}}.
\end{equation}
Introducing a quantum state $\left\vert \psi(t)\right\rangle $ we see that
$\left\langle \psi\right\vert \widehat{\dot{O}}\left\vert \psi\right\rangle
=\frac{d}{dt}\left\langle \psi\right\vert \hat{O}\left\vert \psi\right\rangle
$ only if $\left\vert \psi(t)\right\rangle $ satisfies the equation%
\begin{equation}
\hat{\omega}\left\vert \psi(t)\right\rangle =0 \label{GSschro}%
\end{equation}
with $\hat{\omega}=-i\hbar d/dt+\hat{H}$. Eq.\ $\left(  \ref{GSschro}\right)
$ is a generalization of the Schr\"{o}dinger equation and is the quantum
evolution for all quantum systems including singular non-autonomous ones. The
decomposition $\hat{\omega}=-i\hbar d/dt+H$ is not always valid, as for
instance, for a relativistic Lagrangian $H=0$ (see later on), but in general
we can write $\hat{\omega}=-i\hbar d/dt+\hat{H}_{eff}(t).$ As a first check,
consider the regular Lagrangian $L=\frac{m}{2}\mathbf{\dot{r}}^{2}%
-U(\mathbf{r},t)$ that we write $\tilde{L}d\tau=\mathbf{pdr}-Hdt-td\omega
-\omega d\tau$ with $H=\mathbf{p}^{2}/2m+U(\mathbf{r},t)$. From Eq. $\left(
\ref{bracket}\right)  $ we get the canonical relations $\left[  \hat{x}%
_{i},\hat{p}_{j}\right]  =i\hbar\delta_{ij}$, and the commutators $\left[
\mathbf{\hat{r}},\hat{\omega}\right]  =i\hbar\mathbf{\hat{p}}/m=i\hbar
\widehat{\mathbf{\dot{r}}}$ and $\left[  \mathbf{\hat{p},}\hat{\omega}\right]
=-i\hbar\mathbf{\nabla}\hat{U}=i\hbar\widehat{\mathbf{\dot{p}}}.$ These
commutators are satisfied for $\hat{\omega}=-i\hbar d/dt+\hat{H}$ as expected.\ 

Note that here the time operator does commute with the Hamiltonian $[\hat
{t},\hat{H}]=0$ but instead $[\hat{t},\hat{\varepsilon}]=i\hbar.$ This is
physically correct as in quantum physics the energy can be measured with
arbitrary precision at any time. The common idea that a time operator has to
satisfy $[\hat{t},\hat{H}]=i\hbar$, comes from the usually assumed relation
$d\hat{t}/d\tau=\frac{1}{i\hbar}[\hat{t},\hat{H}]=1$ which is wrong (even for
regular systems), instead $d\hat{t}/d\tau=\frac{1}{i\hbar}[\hat{t},\hat
{\omega}]=1.$ Therefore the problem of unboundedness of the energy spectrum
\cite{Pauli} does not exist here.

In the Heisenberg picture $\hat{\xi}_{H}^{i}=U^{-1}\hat{\xi}^{i}U$ with $U(t)$
the time evolution operator $\left\vert \psi(t)\right\rangle =U\left\vert
\psi\right\rangle _{H}.$ Using Eqs. $\left(  \ref{de1}\right)  $ and $\left(
\ref{GSschro}\right)  $ we find
\begin{equation}
\frac{d\hat{\xi}_{H}^{i}}{dt}=\frac{1}{i\hbar}\left[  \hat{\xi}_{H}^{i}%
,\hat{\omega}_{H}\right]  =\left\{  \xi^{i},\omega\right\}  _{\xi=\hat{\xi
}_{H},\omega=\hat{\omega}_{H}}%
\end{equation}
which is exactly given by Eq. $\left(  \ref{EqFJ}\right)  $ with $\xi^{i}$
replaced by $\hat{\xi}_{H}^{i}.$

We will now apply the Faddeev-Jackiw quantization approach to two examples of
non-autonomous systems.

\bigskip

\textit{Damped harmonic oscillator. }Consider the following singular non
autonomous Lagrangian introduced by \cite{Gd}%
\begin{equation}
L=\frac{1}{2}e^{2\alpha t}\left(  y\dot{x}-x\dot{y}-y^{2}-2\alpha
xy-\Omega^{2}x^{2}\right)  .
\end{equation}
In Dirac formulation this Lagrangian describes a singular system with
time-dependent second-class constraints. The Euler-Lagrange equation
corresponds to the damped harmonic oscillator whose quantization has been
treated in \cite{Gd} via the extended Dirac formalism for singular non
autonomous system.\ With the Faddeev-Jackiw approach we start from the
transformed Lagrangian one form%

\begin{equation}
\tilde{L}d\tau=\frac{1}{2}e^{2\alpha t}\left(  ydx-xdy\right)  -Hdt-td\omega
-\omega d\tau, \label{Lanh}%
\end{equation}
where $H=\frac{1}{2}e^{2\alpha t}(y^{2}+2\alpha xy+\Omega^{2}x^{2}).$ The
element of the $2\times2$ antisymmetric matrix $f_{x,y}^{-1}$ is easily
computed and leads to%
\begin{equation}
\left[  \hat{x},\hat{y}\right]  =i\hbar e^{-2\alpha\hat{t}}. \label{XYDH1}%
\end{equation}
From Eq. $\left(  \ref{C4}\right)  $ as usual $[\hat{t},\hat{\omega}]=i\hbar$
and
\begin{align}
\left[  \hat{x},\hat{\omega}\right]   &  =i\hbar\hat{y}\label{WT}\\
\left[  \hat{y},\hat{\omega}\right]   &  =-i\hbar\left(  \Omega^{2}\hat
{x}+2\alpha\hat{y}\right)  .
\end{align}
Note that the whole set of brackets could be obtained at once by computing the
full $4\times4$ matrix $F_{ij}^{-1}$ directly. The commutator Eq. $\left(
\ref{XYDH1}\right)  $ is the same as in \cite{Gd}. In the Schr\"{o}dinger
picture $\hat{t}=t$, $\hat{\varepsilon}=-i\hbar d/dt$ and from Eq.\ $\left(
\ref{de1}\right)  $ we deduce $\hat{x}(t)=\hat{x}(0)e^{-\alpha t}\ $and
$\hat{y}(t)=\hat{y}(0)e^{-\alpha t}$. From Eq. $\left(  \ref{XYDH1}\right)  $
we see that $\hat{x}(0)$ and $\hat{y}(0)$ are conjugate to each
other.\ Therefore, if we consider for instance the position representation we
have $\hat{x}(t)=xe^{-\alpha t}$ and $\hat{y}(t)=-i\hslash e^{-\alpha
t}\partial_{x}$ acting on the wave function $\psi(x,t).$ The commutators
involving $\hat{\omega}$ leads to the natural choice%
\begin{equation}
\hat{\omega}=-i\hslash\partial_{t}-\frac{1}{2}\hslash^{2}\partial_{x}%
^{2}-i\alpha\hslash x\partial_{x}-\frac{i\alpha\hslash}{2}+\frac{\Omega^{2}%
}{2}x^{2}%
\end{equation}
and the Schr\"{o}dinger equation of the damped harmonic oscillator is
$\hat{\omega}\psi(x,t)=0$ which is the same equation than in \cite{Gd}. This
shows the equivalence of the two methods for the quantum damped harmonic oscillator.

\bigskip

\textit{Relativistic Lagrangian.\ }We consider now the quantization of a
relativistic point like particle in an external electromagnetic field.\ This
system meets specific difficulties such as a null Hamiltonian and the presence
of a gauge symmetry due to the arbitrary choice of the time parametrization
\cite{GT,Gr,Gr2} \ We will see that the Faddeev-Jackiw approach for
non-autonomous systems can be straightforwardly applied and will lead to the
Dirac equation. Consider a relativistic particle interacting with an
electromagnetic potential $A^{\mu}=(\phi,\mathbf{A})$ whose quadri-position
$x^{\mu}=x^{\mu}(\tau)$ with $\mu=0..3$ depends on a parameter $\tau$. The
metric is ${\large \eta}_{\mu\nu}=$ diag$(1,-1,-1,-1).$ The reparametrization
invariant action is $S=\int L_{\tau}d\tau$ with a Lagrangian one-form%
\begin{equation}
L_{\tau}d\tau=-mds-eA^{\mu}dx_{\mu} \label{LR}%
\end{equation}
in unit $c=1$, with $ds=\sqrt{dx^{\mu}dx_{\mu}}.$ The quadri-momenta $p_{\mu
}=-mx_{\mu}^{\prime}/s^{\prime}-eA_{\mu}$ lead to a null Hamiltonian $H_{\tau
}=p_{\mu}x^{\prime\mu}-L=0$ as $L$ is homogeneous of degree one in the
velocities. Still we have the constraint%
\begin{equation}
\left(  p_{0}+e\phi\right)  ^{2}-\mathbf{\mathbf{(\mathbf{p}-}}%
e\mathbf{\mathbf{\mathbf{A})}}^{2}=m^{2}.
\end{equation}
By analogy to the historical calculus of Dirac, it can be linearized by
introducing the usual Clifford algebra to get the equivalent constraint%

\begin{equation}
\Lambda=p_{0}+e\phi+\boldsymbol{\alpha}\cdot\mathbf{\mathbf{(\mathbf{p}-}%
}e\mathbf{\mathbf{\mathbf{A)}}}+\beta m=0, \label{Cons}%
\end{equation}
where the Clifford generators $\alpha_{i}$\ ($i=1,2,3$) and $\beta$\ satisfy
the relations $\alpha_{i}\alpha_{j}+\alpha_{j}\alpha_{i}=2\delta_{ij}%
,$\ $\beta^{2}=1$\ and $\alpha_{i}\beta+\beta\alpha_{i}=0.$\ Dirac matrices
are a four-dimensional representation of this Clifford Algebra. The constraint
Eq. $\left(  \ref{Cons}\right)  $ is implemented by a Lagrange multiplier term
$\lambda^{\prime}\Lambda$ added to $L_{\tau}$. The gauge invariance due to the
arbitrary choice of $\tau$ of Eq. $\left(  \ref{LR}\right)  $ is again fixed
by adding $\omega\left(  t^{\prime}-1\right)  d\tau$ to $L_{\tau}$ (a
condition equivalent to $t=\tau$) and we get the one-form Lagrangian%

\begin{equation}
\tilde{L}d\tau=\mathbf{p}d\mathbf{x+}p_{0}dt+\Lambda d\lambda-td\omega-\omega
d\tau. \label{LRelat}%
\end{equation}
It has the same form as Eq.\ $\left(  \ref{L2}\right)  $ if we introduce the
$10$-component phase-space coordinates $\zeta^{i}=(\mathbf{x},t,\mathbf{p}%
,p_{0},\lambda,\omega)$. The matrix \ $F_{ij}^{-1}$ can be easily computed and
written formally as%
\[
F_{ij}^{-1}=\left(
\begin{array}
[c]{cccccc}%
0 & 0 & 1 & -\boldsymbol{\alpha} & 0 & \boldsymbol{\alpha}\\
0 & 0 & 0 & 0 & 0 & 1\\
-1 & 0 & 0 & \mathbf{D} & 0 & -\mathbf{D}\\
\boldsymbol{\alpha} & 0 & -\mathbf{D} & 0 & -1 & -D_{0}\\
0 & 0 & 0 & 1 & 0 & -1\\
-\boldsymbol{\alpha} & -1 & \mathbf{D} & D_{0} & 1 & 0
\end{array}
\right)  ,
\]
where $\mathbf{D}=e\boldsymbol{\nabla}(\phi-\boldsymbol{\alpha}\cdot
\mathbf{\mathbf{\mathbf{A)}}}$ and $D_{0}=e\partial_{t}\left(  \phi
-\boldsymbol{\alpha}\cdot\mathbf{\mathbf{\mathbf{A}}}\right)  .$ From
$F_{ij}^{-1}$ we read off the non-vanishing brackets:%
\begin{align}
\left\{  \mathbf{x,p}\right\}   &  =\mathbf{1},\text{ \ \ }\left\{
t,\omega\right\}  =1\label{XPD1}\\
\left\{  \mathbf{x,}\omega\right\}   &  =\boldsymbol{\alpha}\mathbf{,}\text{
\ \ \ }\left\{  \mathbf{x,}p_{0}\right\}  =-\boldsymbol{\alpha}\label{XPD2}\\
\left\{  \mathbf{p},\omega\right\}   &  =-\mathbf{D,}\text{ \ \ \ }\left\{
\mathbf{p},p_{0}\right\}  =\mathbf{D}\label{XPD3}\\
\left\{  \lambda\mathbf{,}\omega\right\}   &  =-1,\text{ \ \ \ }\left\{
\lambda,p_{0}\right\}  =1\label{XPD4}\\
\left\{  p_{0}\mathbf{,}\omega\right\}   &  =-D_{0}. \label{XPD5}%
\end{align}
The bracket $\left\{  \mathbf{x,p}\right\}  =\mathbf{1}$ justifies Dirac's
choice of the usual commutation relations after the linearization of the
Hamiltonian.\ Looking at the Hamilton equations $d\zeta^{i}/d\tau=\left\{
\zeta^{i},\omega\right\}  $ we first see that $\lambda^{\prime}=-t^{\prime
}=1.$ Thus the variable $\lambda$ is redundant and can be dismissed for the
quantization.\ The brackets involving $p_{0}$ show that despite $H=0,$ it is
$-p_{0}$ that plays the role of an effective Hamiltonian of the initial
system. For the quantum version an obvious choice which satisfies all the
commutators is clearly%
\begin{equation}
\hat{\omega}=-i\hbar\frac{d}{dt}-\hat{p}_{0}.
\end{equation}
Therefore the time evolution of the quantum state given by $\hat{\omega
}\left\vert \psi(t)\right\rangle =0$ leads to%
\begin{equation}
i\hbar\frac{d}{dt}\left\vert \psi(t)\right\rangle =\left(  \boldsymbol{\alpha
}\cdot\mathbf{\mathbf{(\mathbf{\hat{p}}-}}e\mathbf{\hat{A}\mathbf{\mathbf{)}}%
}+\beta m+e\hat{\phi}\right)  \left\vert \psi(t)\right\rangle ,
\end{equation}
which is nothing else but the Dirac equation of a relativistic quantum
particle in an electromagnetic field.\ This achieves our goal of deriving the
relativistic Schr\"{o}dinger equation, i.e., the Dirac equation from the
canonical quantization of the classical relativistic Lagrangian. This
quantization method could be generalized to a particle moving in a curved
space and more generally to the case of time invariant reparametrization systems.

\bigskip

\textit{Conclusion}.\ An extension of the Faddeev-Jackiw method in order to
solve the problem of time dependent constraints has been considered. For that
purpose a time parameter is introduced to treat the time as a dynamically
variable, which is accompanied by the emergence of gauge symmetry. This one is
fixed with the help of a supplementary variable that plays the role of a
conserved Hamiltonian. After obtaining the correct brackets, we were able to
give the most general form of the quantum (Schr\"{o}dinger) equation, valid
also for singular non-autonomous systems. The method can be naturally applied
to the case of a relativistic particle\ in an external electromagnetic
field.\ The theory developed in this paper should be useful for the
quantization of physical constrained systems in the presence of time-dependent
external fields.

\end{document}